\documentclass{article}

\usepackage[final]{neurips_2023}

\usepackage[utf8]{inputenc} 
\usepackage[T1]{fontenc}    
\usepackage{hyperref}       
\usepackage{url}            
\usepackage{booktabs}       
\usepackage{amsfonts}       
\usepackage{nicefrac}       
\usepackage{xcolor}         
\usepackage{graphicx}
\usepackage{float}
\usepackage{wrapfig}

\graphicspath{ {./images/} }

\title{Emergence of Collective Open-Ended Exploration from Decentralized Meta-Reinforcement Learning}

\author{
  Richard Bornemann \thanks{Equal first authors} \thanks{Work done as an intern at Flowers} \\
  Inria - Flowers Team \\
  \texttt{richard.bornemann@inria.fr}
  \And
  Gautier Hamon \footnotemark[1] \\
  Inria - Flowers Team \\
  \texttt{gautier.hamon@inria.fr}
  \AND
  Eleni Nisioti \\
  Inria - Flowers Team
  \And
  Clément Moulin-Frier \\
  Inria - Flowers Team 
}

\begin{document}

\maketitle

\begin{abstract}
Recent works have proven that intricate cooperative behaviors can emerge in agents trained using meta reinforcement learning on open ended task distributions using self-play. While the results are impressive, we argue that self-play and other centralized training techniques do not accurately reflect how general collective exploration strategies emerge in the natural world: through decentralized training and over an open-ended distribution of tasks. In this work we therefore investigate the emergence of collective exploration strategies, where several agents meta-learn independent recurrent policies on an open ended distribution of tasks. To this end we introduce a novel environment with an open ended procedurally generated task space which dynamically combines multiple subtasks sampled from five diverse task types to form a vast distribution of task trees. We show that decentralized agents trained in our environment exhibit strong generalization abilities when confronted with novel objects at test time. Additionally, despite never being forced to cooperate during training the agents learn collective exploration strategies which allow them to solve novel tasks never encountered during training. We further find that the agents learned collective exploration strategies extend to an open ended task setting, allowing them to solve task trees of twice the depth compared to the ones seen during training. Our open source code as well as videos of the agents can be found on \href{https://sites.google.com/view/collective-open-ended-explore}{our companion website}

\end{abstract}

\section{Introduction}

Cooperative exploration plays a pivotal role in fostering the collective intelligence in groups of autonomous agents. Developing strategies to effectively coordinate the exploration of large search spaces has the potential to significantly decrease the time needed to find optimal solutions. The power of cooperative exploration can be seen in areas ranging from complex search and rescue missions to the entire field of modern science, where scientists work together to coordinate their research. Studying the emergence of cooperative behavior in artificial agents has therefore garnered much interest, especially in the field of multi agent reinforcement learning \cite{panait2005cooperative} \cite{tan1993multi} \cite{ishiwaka2003approach}. With the recent successes of deep reinforcement learning, the difficulty of the tasks being researched have significantly increased leading to a corresponding increase in the complexity of learned cooperative behaviors \cite{berner2019dota} \cite{jaderberg2019human}. Extending multi agent reinforcement learning further by training agents on open ended task spaces has lead to agents which exhibit strong generalization abilities, being able to adapt to novel tasks through strong exploration priors acquired during training \cite{openendedlearningteam2021openended} \cite{team2023human}. 

However, these works do not study the simultaneous training of decentralized agents, but rather make use of techniques such as self play or playing against static checkpoints of other agents. We argue that this approach does not accurately reflect how autonomous agents learn together in the real world. Rather, when confronted with novel tasks in a group setting, all autonomous agents in the group are exploring and actively updating their prior believes, forcing them to either explicitly or implicitly coordinate their learning progress to converge to some shared strategy which allows them to effectively solve the task. Recent works such as\cite{de2020independent} \cite{liu2019emergent} \cite{perolat2017multi} have shown that this coordination process can emerge in decentralized multi agent reinforcement learning, leading to independent agents learning to solve complex tasks together. 

In this work we want to further investigate the emergence of  cooperative exploration strategies of decentralized agents by training them on a open ended distribution of tasks. To this end we introduce a novel environment which is conceptually simple yet allows for a complex open ended procedurally generated task space by dynamically combining multiple subtasks sampled from five task types to form a task tree which needs to be solved sequentially (Fig.~\ref{fig:test-sampling}), akin to the notion of recipes in \cite{team2023human}. We train two agents parametarized by independent recurrent neural networks and optimized using standard proximal policy optimization. As no information is given to the agents about which subtasks have been sampled or how and in which order they should be solved, the agents have to develop general strategies for exploring the environment, effectively learning how to learn from the information obtained by interacting with the environment throughout the episode, in order to solve novel tasks. We show that training independent decentralized agents on only multi agent episodes leads to sub-optimal behavior of the agents, primarily due to the problem of credit assignment when rewards are shared between agents. We propose to include single agent episodes during training to force the agents to learn to solve tasks on their own without relying on any help from other agents. We find that training on a mixture of single and multi agent episodes increases the agents individual performance while simultaneously decreasing the individual performance differences between the agents, leading to a strong improvement in performance in multi agent tasks. 

Using this approach we find that decentralized agents trained in our environment learn a powerful collective exploration strategy, allowing them to solve over $70$ percent of task trees encountered during training. Moreover, these powerful exploration capabilities lead to strong generalization performance when confronted with objects unseen during training, as well as on novel tasks which require complex coordination to be solved successfully at test time. Additionally we show that the learned collective exploration strategies extend to the open ended task setting, enabling the agents to effectively generalize to task trees with a depth of six, featuring an increased complexity of subtasks, despite being initially trained on task trees comprising only three subtasks.

\section{Related Work}
Cooperative behavior in multi-agent environments has long been a topic of great interest in reinforcement learning \cite{panait2005cooperative} \cite{tan1993multi} \cite{ishiwaka2003approach}. Recently, techniques from multi-agent reinforcement learning such as self-play have been used to train agents to human level performance in areas ranging from board games \cite{silver2016mastering} to complex modern video games \cite{berner2019dota} \cite{vinyals2019grandmaster}. Other works study the emergence of coordination and cooperation in populations of agents in complex competitive environments. \cite{baker2020emergent} has shown that teams of agents competing against each other in the game hide and seek can develop sophisticated strategies such as tool use and even learn to exploit bugs in the environments implementation. These works make use of forms of centralized training, such as shared agent parameters or self-play to achieve their impressive results. \cite{jaderberg2019human} and \cite{liu2019emergent} have shown that decentralized methods, when combined with population based training, can lead to the emergence of complex shared cooperation and coordination strategy within teams of agents. \cite{de2020independent} have further shown that a simplified approach of training independent agents without any centralized information sharing or population based training can lead to competitive performance on the Starcraft Multi Agent Challenge \cite{samvelyan2019starcraft}.

In order for agents to deal with environments where the task at hand is unknown to the agent and sampled from large distribution of possible tasks, meta-learning has been proposed. Meta-learning allows an agent to learn to use its existing knowledge to quickly adapt to new tasks at test time \cite{schmidhuber1996simple}. Combining this approach with reinforcement learning and recurrent neural networks has lead to agents that are able meta-learn their own reinforcement learning algorithm, allowing them to adapt and solve novel tasks \cite{wang2017learning} \cite{duan2016rl2}. Recent works have shown such Meta Reinforcement Learning (Meta RL) algorithms to be very effective resulting in multi task robots that are able to adapt to new tasks \cite{yu2020meta} \cite{nagabandi2018learning} and environments \cite{akkaya2019solving} on the fly, even allowing them to generalize their behaviors from simulations to the real world. \cite{justesen2018illuminating} shows that combining Meta RL with open ended procedurally generated environments facilitates open ended skill acquisition and allows agents to better adapt to novel environments. Similarly \cite{openendedlearningteam2021openended} and \cite{team2023human} show that agents trained on vast diverse task spaces are able to quickly adapt to novel tasks, even surpassing human adaptation skills.

Work in combining multi-agent environments with Meta RL has so far remained relatively sparse. \cite{parisotto2019concurrent} are exploring the use of multiple agents acting simultaneously to efficiently explore the environment in order to then leverage their pooled knowledge in the exploitation phase to solve complex tasks. \cite{samvelyan2023maestro} uses Meta RL together with open ended environment design to train a pool of agents on competitive two player tasks. Closer to our work \cite{openendedlearningteam2021openended} and \cite{team2023human} also present results for agents trained in multi agent settings in a multi task and Meta RL fashion with open ended task distributions. However, these methods employ either static checkpoints from a population of agents or older versions of themselves to train an agent in multi agent episodes. Our work therefore differs from this approach by training two decentralized agents together in the same environment, without making use of techniques like self-play or population based training, commonly used in other works on emergent cooperation. 
\label{rel_work}

\section{Method}
\begin{figure}
    \centering
    \includegraphics[width=\textwidth]{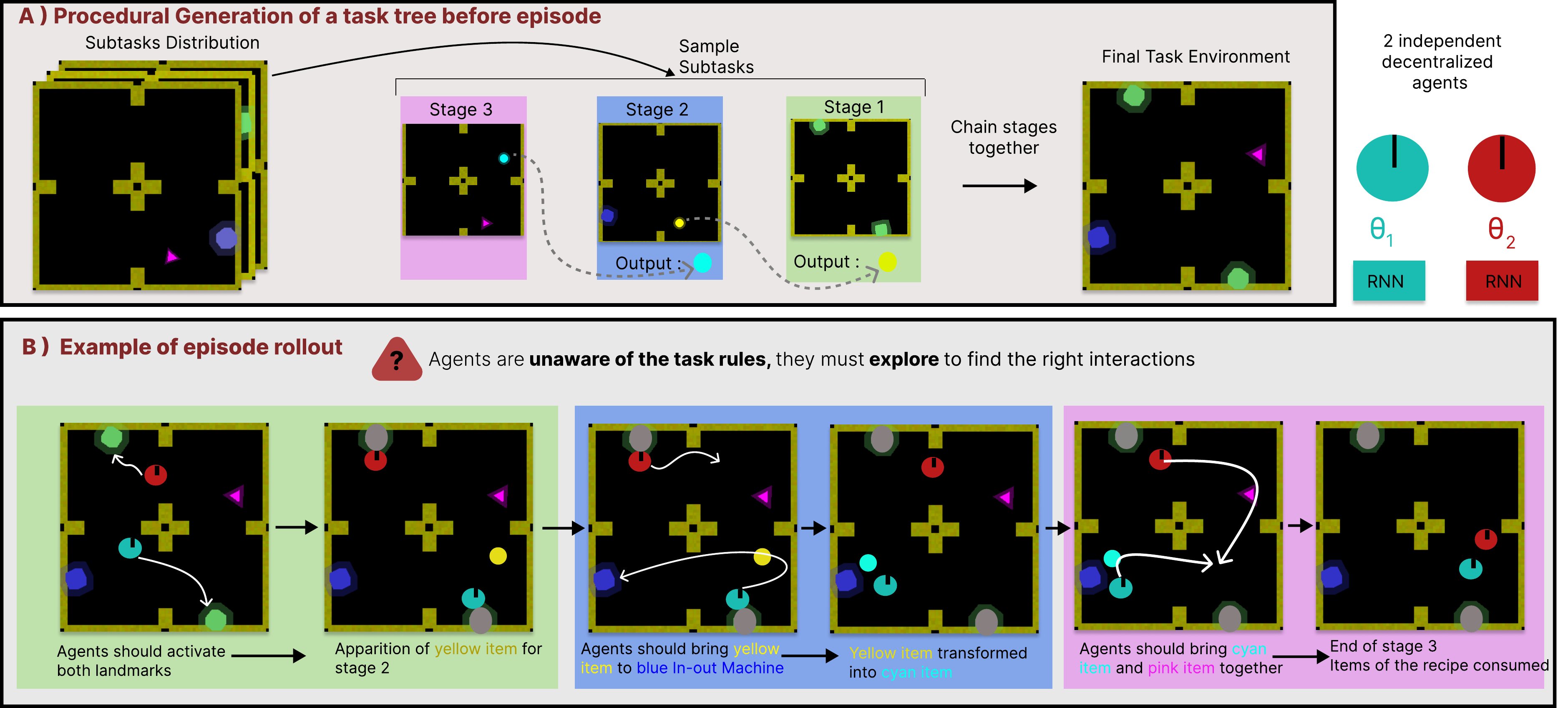}
    \caption{\textbf{Task Tree Sampling and Episode Rollout.} \textbf{A)} shows the task tree sampling process. First three subtasks are sampled from the distribution of subtasks (Section~\ref{task_types}), one for each stage of the task tree. All of the objects required to solve the subtask for stage one and some of the objects required by subtasks in later stages are then placed in the environment. The remaining objects required to solve the later subtasks can be created through solving preceding subtasks (Section ~\ref{task_sampling}). \textbf{B)} shows an example of a single episode rollout. The agents have to complete the subtasks sequentially in order to create objects which are needed by the subtasks in later stages. Since a new task tree with different subtasks is sampled at the beginning of each episode and no information about the subtasks is given to the agents, the agents have to explore the environment and interact with all present objects so solve the subtask at each stage. Videos of the agents behaviors can be found on \href{https://sites.google.com/view/collective-open-ended-explore}{our companion website}.  }
    \label{fig:test-sampling}
\end{figure}

\subsection{Environment}
Our 2D environment is implemented using Simple-Playgrounds \cite{Simple-Playgrounds} and features realistic physics for object movements and collisions, as well as a range of interaction dynamics for different object types. The agents have two continuous movement actions for turning angles and forward walking speeds, as well as the two discrete grasping and activating actions for interacting with the objects present in the environment. The agents are able to pass through each other and cannot grasp an object which is currently being held by the other agent. This is done in order to limit noise caused by the agents interfering with each other during training. As input the agents get a limited top down view of their surroundings, preventing them from having full vision of the environment. The environment itself consists of rooms connected by large doorways, preventing the agents from diagonally crossing the map (Fig.~\ref{fig:test-sampling}). We differentiate between two object types required by the subtasks. Environment objects such as landmarks are large square shaped immovable objects that are spawned at the edges of the environment. Task objects are smaller and can be moved by the agents. Objects always have some form of interaction dynamic either with another task object, an environment object or an agent. The different interaction dynamics of environment objects, task objects and agents are explained in detail in (Section~\ref{task_types}). The pool of possible objects includes three different shapes and colors for a total of nine different task objects, as well as a further four different environment objects. Task objects are always randomly sampled before each episode and do not poses any fixed interaction dynamic, encouraging the agents to explore by trying to combine different task objects together to create new objects. Environment objects however always possess the same interaction dynamic. Agents should therefore meta-learn how they can interact with a specific environment object.

\subsubsection{Task types}
\label{task_types}
We include five different subtask types in our environment. Although all of the tasks can be solved by a single agent alone, the agents should nonetheless learn to cooperate in order to solve the tasks as quickly and efficiently as possible.

\textbf{Activate Landmarks:} The agents are tasked with locating one or two landmarks, which are randomly placed at the edges of the environment, and activate them. In the two landmark case, the agents have to activate both of the landmarks within three hundred environment steps of activating the first landmark. The agents are expected to learn to split up and independently locate a landmark in order to solve the task as quickly as possible.

\textbf{Lemon Hunt:} The agents need to find a specific object and switch into the "lemon" object by activating it. The resulting lemon object can then be consumed by either agent. The agents should learn to interact with all the task objects present in the environment and try to activate them.

\textbf{Crafting:} The agents need to combine two objects in the environment to either spawn a new object or make an existing object disappear. The agents are expected to first explore the environment until finding a task object of interest and trying to combine it with other task objects. Additionally the agents should coordinate to bring task objects together and not explore object combinations which have already been tried by the other agent.

\textbf{In Out Machine:} The agents need to find the correct object and bring it to the in out machine, which is randomly spawned at the edge of the environment. The object is then switched into an object required by following tasks. Therefore to efficiently solve this task, the agents should try bringing all the objects present in the environment to the in out machine until they find the correct object which can be switched.

\textbf{Drop Off Point:} Similar to the in out machine, only now the agents need to bring the correct object to the drop off point to make it disappear of completing all of the preceding subtasks.

\subsubsection{Procedural Generation of Task Trees}
\label{task_sampling}
At the beginning of each episode, $d$ subtasks are selected from categories of five different task types in order to build a task tree of depth $d$ (Fig.~\ref{fig:test-sampling}), similar to the concept of task recipes in \cite{team2023human}. Initially the end condition that must be satisfied for the last task in the task tree to be considered a success is first selected, with the possible end conditions being \textbf{object exists} or \textbf{object does not exist}. The subtasks are then sampled recursively to depth $d$, where the subtask at each level outputs the objects required by the subsequent task. The set of subtasks which can be sampled at each stage depends on the stage and the subtasks sampled in preceding stages. At each stage, the objects required by the subtask are sampled uniformly from a pool of nine different objects. The environment objects required by all sampled subtasks and the objects required for the first subtask are then spawned in at random points in the environment. Using this method, we can build complex task trees with arbitrary depth, procedurally generating an open-ended distribution of tasks.

\subsection{Training Setup}
\subsubsection{Reward Structure}
 After successfully completing a subtask in the task tree agents are jointly rewarded for each time step until the end of the episode, encouraging them to continue improving their performance even in subtasks with very high success rates. Additionally, the reward per time step increases exponentially for completing subtasks in higher stages, incentivizing the agents to learn to better solve subtask of higher stages rather then marginally improving their performance in the easier to solve lower stages. We observe that this reward structure significantly improves performance over rewarding the successful completion of each stage the same. The reward structure in combination with the tasks trees made up of different subtasks leads to a smooth improvement of the agent across different stages during training, eliminating the need for any explicit form of curriculum design.

\subsubsection{Agent Architecture}
We employ a similar approach to \cite{de2020independent}, where each agent in the environment is independently parameterized by a neural network which is optimized using standard proximal policy optimization \cite{schulman2017proximal}, without sharing any parts of the network. Each agent only gets as input its own limited top down view of the environment as well as its own action and reward from the previous step as usually done in Meta-RL. No additional information about the task tree, environment objects or the other agent are given to the agents. For the agent architecture we use the same convolutional neural network used in \cite{Mnih2015HumanlevelCT}, followed by a one layer fully connected network of size [256] whose output is fed through a ReLU non-linearity and concatenated with the agents action and reward from the previous step. This is followed by a one layer LSTM \cite{6795963} with size [256] and finally by a policy head consisting of three layers with sizes [64, 64, 4] and a value head with sizes [64, 64, 1], where each of the hidden layers is followed by a ReLU non-linearity. 

\subsubsection{Agent Training}
At the beginning of each episode, we first sample whether the episode will be played in the multi agent or single agent paradigm. In the multi agent case, the agents will be placed in the same environment, whereas in the single agent case the agents will play in two different environments without interacting with each other. We then sample one task tree in the multi agent case or two task trees, one for each environment, in the single agent case through the procedure described in (Section~\ref{task_sampling}). Since all subtasks can be solved by a single agent during training, we do not modify the task distribution for single agent episodes. As is common in Meta-RL, the agents do not get any information about which subtasks have been sampled and in which order they should be solved. They are then randomly placed in the environment and have a limit of $1000$ environment steps to solve all the subtasks sequentially. After the limit is up, the environment is reset, we resample the multi or single agent setting and sample one or two new task trees. We train on batches of $480$ complete episodes for a total of $750000$ episodes.
We linearly decay the learning rate to $0$ from a starting value of $0.00025$ over the course of training. As the agents are fully decentralized, no information is shared between them during training. We therefore update the parameters of each agents network based only on its own experiences from each batch. 

\section{Results}
In this section, we present the experimental results for our training paradigms, both with and without single-agent episodes. We define the performance of our agents as the rate of completed subtasks per stage. We find that including single agent episodes improves the performance during training. Further, we show that decentralized agents trained in our environment exhibit strong generalization abilities when encountering objects unseen during training and complex novel tasks which require efficient coordination between the agents in order to be solved. Finally, we highlight the agents proficiency in generalizing to the open-ended task setting. This is demonstrated through an evaluation of their performance on task trees encompassing six stages, instead of the depth of three stages encountered during training.

\subsection{Training Performance}
\begin{wrapfigure}{r}{0.4\textwidth}
    \begin{center}
        \includegraphics[width=1.0\linewidth]{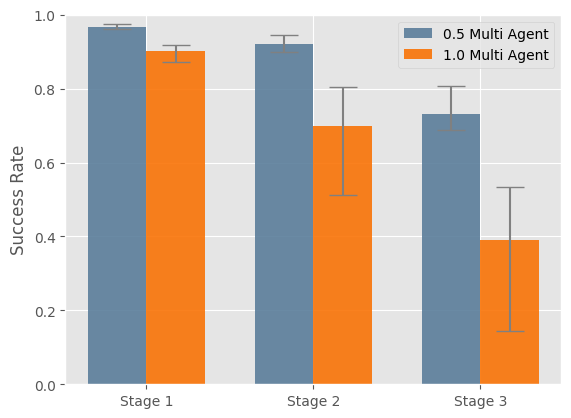}
    \end{center}
    \caption{Success rates for $100$\% vs $50$\% multi agent episodes during training}
    \label{fig:success_rate}
\end{wrapfigure}

\begin{figure}[H]
  \centering
  \begin{minipage}[b]{0.4\textwidth}
    \includegraphics[width=\textwidth]{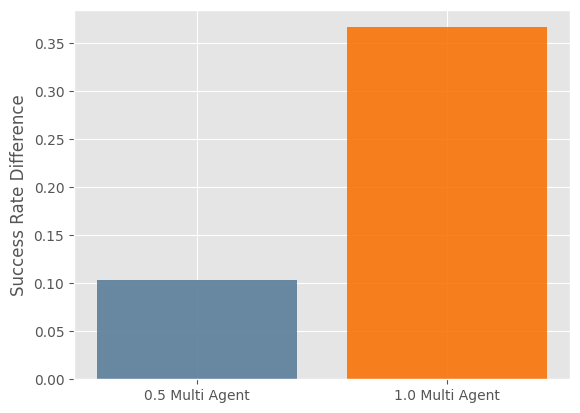}
    \caption{Mean stage $3$ success rate difference between two agents trained on $100$\% vs $50$\% multi agent episodes}
    \label{fig:difference}
  \end{minipage}
  \hfill
  \begin{minipage}[b]{0.4\textwidth}
    \includegraphics[width=\textwidth]{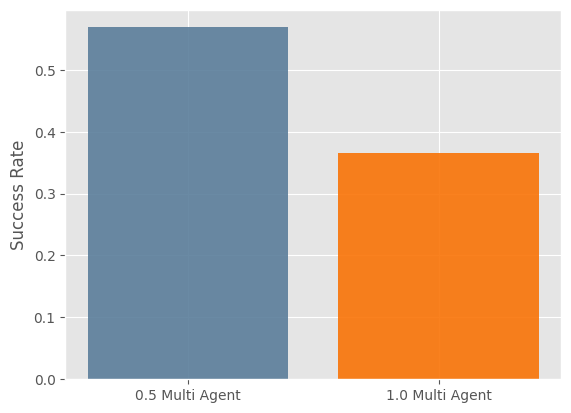}
    \caption{Mean stage $3$ success rate on single agent episodes for agents trained on $100$\% vs $50$\% multi agent episodes}
    \label{fig:single success}
  \end{minipage}
\end{figure}

Our findings indicate that the straightforward approach of training two decentralized agents together solely on multi-agent episodes results in suboptimal performance. When looking at the individual performances of the agents during single-agent episodes, we observe significant disparities in skill levels (Fig.~\ref{fig:difference}). We argue that these performance discrepancies stem from the credit assignment problem, which emerges due to multiple agents sharing rewards \cite{rahaie2016critic}. When one agent accomplishes a subtask, both agents receive the reward, resulting in potentially misleading parameter updates for the agent that was not directly involved in completing the subtask. This dynamic can magnify minor skill disparities at the outset of training, ultimately culminating in substantial differences in learned behaviors by the end of training. To combat this problem we propose training the agents on both single and multi agent episodes, as the multi-agent credit assignment problem can not arise during single agent episodes. We find that this greatly decreases the skill differences between the agents and increases the single agent performances (Fig.~\ref{fig:single success}), leading to a large gain in performance in multi agent episodes. While agents trained on multi and single agent episodes are able to solve the vast majority of subtasks for all stages within the time limit, agents trained solely on multi agent episodes perform worse on stages one and two and fail to solve subtasks in stage three in the majority of episodes. The large dropoff in success rate from stage two to stage three is mainly caused by the agents running out of time. This indicates that agents trained on multi and single agent episodes are able to solve the subtasks much quicker when compared with agents that where only trained on multi agent episodes. In the following we therefore limit our evaluations to the case of mixed single and multi agent training.

\vspace{1.0 cm}

\subsection{Generalization Performance}
\subsubsection{Novel Objects}
\begin{wrapfigure}{r}{0.4\textwidth}
    \begin{center}
        \includegraphics[width=1.0\linewidth]{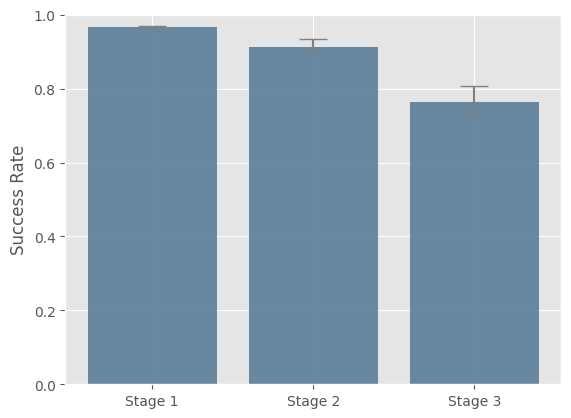}
    \end{center}
    \caption{Success rates with novel objects}
    \label{fig:success_rate_novel}
\end{wrapfigure}

We replace all task objects present during training with novel shapes and colors and evaluate the agents performance on the training task distribution in multi agent episodes. We find that including novel task objects does not lead to any decrease in performance when compared to the training performance with the standard task objects (Fig.~\ref{fig:success_rate_novel}). When looking at videos of the agents we observe that they interact with novel task objects in the same fashion as they would with task objects seen during training. They explore the possible task object combinations to find environment objects or other task objects which lead to a successful interaction. As the colors and shapes of task objects do not carry any information about the task objects interaction possibilities, we suspect that the agents learn to not focus on any specific characteristics of the task objects. Instead, the agents rely on powerful exploration to try out all possible object interactions in the environment to solve tasks, allowing them to generalize seamlessly to novel objects.

\subsubsection{Forced Cooperation}
\label{forced_coop}
\begin{figure}[H]
  \centering
  \begin{minipage}[b]{0.4\textwidth}
    \includegraphics[width=\textwidth]{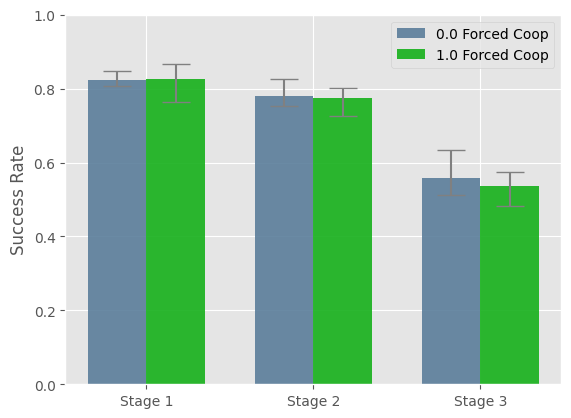}
    \caption{Evaluation for agents trained on $0$\% vs $100$\% forced cooperation tasks}
    \label{fig:forced_coop_eval}
  \end{minipage}
  \hfill
  \begin{minipage}[b]{0.4\textwidth}
    \includegraphics[width=\textwidth]{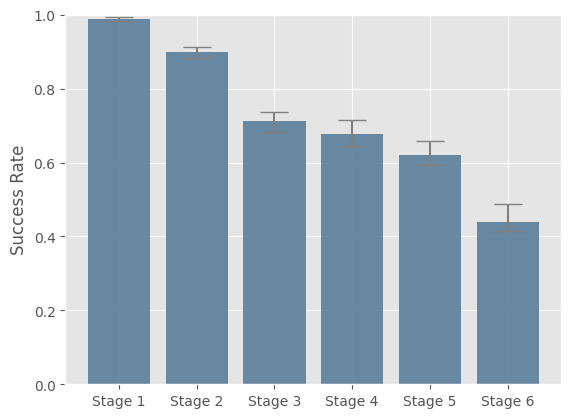}
    \caption{Success rates for open ended exploration on $6$ stages and $4000$ timesteps}
    \label{fig:0.0_6_stage_eval.png}
  \end{minipage}
\end{figure}

To test the ability of the agents to effectively cooperate, we devise three subtasks which require the agents to cooperate, based on the subtasks presented in (Section~\ref{task_types}). We call these tasks "forced cooperative". Detailed descriptions for the forced cooperation subtasks can be found in the Appendix~\ref{appendix:forced_coop}. During evaluation we switch out the landmarks and lemon hunt subtasks for these three forced cooperation subtasks and present the average results over $4800$ episodes. We observe that the agents still show strong performance when being forced to cooperate in order to solve tasks, even with the tasks now requiring much more intricate coordination for successful completion (Fig.~\ref{fig:forced_coop_eval}). This suggests that the cooperative behaviors learned during training are able to help the agents generalize to settings where cooperation is required, despite never encountering such as scenario during training. Notably, the agents trained without any forced cooperative tasks narrowly outperform the agents trained only in the forced cooperative setting, when evaluated on forced cooperative tasks. We hypothesize that this is due to the high difficulty of the forced cooperation tasks preventing the agents to efficiently learn without any form of curriculum. When analyzing the agents behavior, we observe that when confronted with novel behaviors of previously seen environment objects like landmarks, they try to exploit the behaviors learned during training. However, after some amount of tries, the agents realize that their current approach is not working and start exploring the environment until finding an environment or task object which they are able to interact with. Paired with the agents behavior to often first explore the environment separately, they are able to solve complex coordination tasks like the forced cooperation version of the landmarks subtask, where the agents have to find and activate their respective version of the landmark within ten environment steps of each other. To gain a better understanding of how the agents are able to solve forced cooperation tasks where refer the reader to the accompanying \href{https://sites.google.com/view/collective-open-ended-explore}{website with videos} 

\subsubsection{Novel Task}
To further evaluate the agents cooperative abilities and their capacity to generalize to novel tasks we introduce the "pressure plate" task, which the agents have never seen during training. In this task, one agent is tasked with remaining in proximity to the pressure plate landmark and continuously activating it. The second agent is responsible for locating a task object and transporting it to the "in and out" machine, which only works if the pressure plate is activated, in order to switch it to the condition object which indicates successful completion of the task. This task requires coordination between the agents to determine which one remains at the pressure plate and which one explores the environment. We find that the agents are able to solve this task on $45$ percent of the trials, showing impressive generalization abilities. Observing videos of the agents playing this task, we notice a recurring pattern wherein the agents activate the pressure plate and promptly move away. Once the pressure plate deactivates, the agents quickly return to reactivate it, resulting in a repetitive loop where the agents continuously trigger the pressure plate. While both agents frequently become stuck in this loop initially, we observe in many instances that one agent manages to break free and begins to explore the environment. Ultimately, this agent successfully transports the task object to the "in and out" machine, thereby effectively solving the pressure plate task. We hypothesise that the behavior for one agent to break the loop arises during training when agents learn that they should split up their efforts in order to efficiently explore the environment and solve tasks as quickly as possible, similar to the behaviors observed in (Section~\ref{forced_coop}). The agents ability to generalize to novel forced cooperation tasks therefore seems to mainly stem from their exploration and coordination abilities.

\subsubsection{Open Ended Exploration}
We evaluate the agents ability to open endedly explore their environment by setting the number of stages in the task tree to six and increasing the time limit to $4000$ environment steps (Fig.~\ref{fig:0.0_6_stage_eval.png}). We further set the rewards for completing each subtask to zero in order to prevent giving the agents a reward feedback for stages higher than three, which they have not seen during training. We observe that this boosts performance on higher stages during evaluation. We find that agents generalize surprisingly well to task trees with six stages. It is worth emphasizing that as we increase the number of stages in the task tree, the number of task objects present in the environment also increases. Consequently, this exponentially expands the number of possible combinations of task objects that the agents must experiment with in order to solve the subtasks. The agents strong performance therefore not only shows their capacity for collective open ended exploration but also showcases their capacity to tackle subtasks of higher complexity compared to those encountered during training. 

\section{Conclusion}
In this work we investigate the emergence of collective exploration strategies in decentralized agents trained on an open ended distribution of tasks in a Meta RL fashion. While previous related works have studied how cooperative behaviors can emerge from Meta-RL in an open-ended task space \cite{openendedlearningteam2021openended}\cite{team2023human}, our approach is, to our knowledge, the first attempt at demonstrating it in a decentralized training paradigm, together with available \href{https://sites.google.com/view/collective-open-ended-explore}{open source} code for reproducibility. We show that decentralized agents trained only on multi agent episodes exhibit subpar performance and propose to incorporate single agent episodes to boost the individual agents performance. Agents trained using this approach exhibit strong generalization abilities to unseen objects and tasks requiring the agents to cooperate, indicating the emergence of collective exploration strategies despite never being forced to cooperate during training. We further show that the agents are able to generalize their exploration behavior to an open ended setting and solve task trees of twice the length compared to task trees seen during training. We observe that withholding any reward feedback from the agents at test time boosts the exploration performance, suggesting the emergence of an intrinsic motivation to open endedly explore the environment, even in the absence of extrinsic rewards. However, directly isolating which learned behaviors allow the agents to cooperate and coordinate their movements remains difficult. Adding a direct communication channel between the agents could therefore present a promising method to boost the agents multi agent performance and allow for a clearer understanding of their learned cooperative behaviors by analyzing the learned communication. Finally, incorporating more sophisticated approaches to solving the multi agent credit assignment problem and investigating how to boost the agents individual performance while preserving their ability to cooperate holds potential to greatly increase the complexity of the agents learned cooperative behaviors.

\textbf{Acknowledgements} This research was funded by the French National Research Agency (\url{https://anr.fr/}, project ECOCURL, Grant ANR-20-CE23-0006. This work also benefited from access to the Jean Zay (Idris) supercomputer associated with the Genci grant A0151011996.

\bibliographystyle{plainnat}
\bibliography{bibliography}

\appendix

\section{Forced Cooperation}
\label{appendix:forced_coop}
We define a task tree as forced cooperative if at least one of the sampled subtasks requires both agents to solve it. We describe the three forced subtask types below:

\textbf{Activate Landmarks:} Now always two landmarks are randomly spawned at the edges of the environment. Agents are randomly assigned one of the two landmarks which they are able to activate. Additionally, the landmarks now have to be activated within ten environment steps of each other.

\textbf{Meeting Point:} One landmark is randomly spawned at the edges of the environment. The agents both have to be at the landmark and activate it within ten environment steps of each other.

\textbf{Lemon Hunt:} Now one agent is able to switch a specific object into the lemon object while the other agent is able to consume it.

\end{document}